\documentclass[conference]{IEEEtran}
\IEEEoverridecommandlockouts

\usepackage{cite}
\usepackage{amsmath,amssymb,amsfonts}
\usepackage{algorithmic}
\usepackage{graphicx}
\usepackage{textcomp}
\usepackage{xcolor}
\usepackage{subfigure}
\usepackage{booktabs}
\usepackage{array}
\usepackage{placeins}
\usepackage{url}
\begin{document}
\setcounter{topnumber}{6}    
\setcounter{totalnumber}{6}  
\title{FlexMem: High-Parallel Near-Memory Architecture for Flexible Dataflow in Fully Homomorphic Encryption\\
}

\author{
    Shangyi Shi\textsuperscript{1,2,6}, Husheng Han\textsuperscript{1,2,6}, Jianan Mu\textsuperscript{1,2,6}, Xinyao Zheng\textsuperscript{1,2,6}, \\Ling Liang\textsuperscript{3}, Hang Lu\textsuperscript{1,2,4}, Zidong Du\textsuperscript{1,5}, Xiaowei Li\textsuperscript{1,2,7}, Xing Hu\textsuperscript{1,5}\textsuperscript{*}, Qi Guo\textsuperscript{1}\\
    \textsuperscript{1}\textit{SKLP, Institute of Computing Technology, Chinese Academy of Sciences} \\
    \textsuperscript{2}\textit{University of Chinese Academy of Sciences} \quad
    \textsuperscript{3}\textit{Peking University} \quad
    \textsuperscript{4}\textit{ZGC LAB} \\
    \textsuperscript{5}\textit{Shanghai Innovation Center for Processor Technologies, SHIC} \\
    \textsuperscript{6}\textit{Cambricon Technologies} \quad \textsuperscript{7}\textit{CASTEST} \\
    
    \{shishangyi22s, mujianan, hanhusheng20z, zhengxinyao22s\}@ict.ac.cn \\ lingliang@pku.edu.cn \quad  \{luhang, duzidong, lxw, huxing, guoqi\}@ict.ac.cn \\
}

\maketitle

\begin{abstract}
Fully Homomorphic Encryption (FHE) imposes substantial memory bandwidth demands, presenting significant challenges for efficient hardware acceleration. Near-memory Processing (NMP) has emerged as a promising architectural solution to alleviate the memory bottleneck. However, the irregular memory access patterns and flexible dataflows inherent to FHE limit the effectiveness of existing NMP accelerators, which fail to fully utilize the available near-memory bandwidth. In this work, we propose FlexMem, a near-memory accelerator featuring high-parallel computational units with varying memory access strides and interconnect topologies to effectively handle irregular memory access patterns. Furthermore, we design polynomial and ciphertext-level dataflows to efficiently utilize near-memory bandwidth under varying degrees of polynomial parallelism and enhance parallel performance. Experimental results demonstrate that FlexMem achieves $1.12\times$ performance improvement over state-of-the-art near-memory architectures, with $95.7\%$ of near-memory bandwidth utilization.

\end{abstract}


\section{Introduction}\label{sec:introduction}
Fully Homomorphic Encryption (FHE) is regarded as the holy grail of cryptography for allowing arbitrary evaluations to be performed directly on ciphertexts, which is promising in secure outsourced computation, such as privacy-preserving inference~\cite{HEResNet,HELR}, private information retrieval~\cite{gentryPIR, spiral}, and encrypted databases~\cite{he3db}. 
FHE, including TFHE- and CKKS-like schemes, operates on ciphertexts that are $10^3\times\sim10^6\times$ the size of plaintexts and performs complex operations with gigabytes of auxiliary data, resulting in substantial memory and computational overhead. 

Existing FHE accelerators~\cite{craterlake,ARK,SHARP, MAD} have identified the memory bottleneck in CKKS and introduced memory-aware dataflow optimizations~\cite{MAD,ARK}. CKKS accelerators are commonly designed with large on-chip SRAM (180 MB~\cite{SHARP}, 256 MB~\cite{Trinity, craterlake}, 512 MB~\cite{ARK}) for more on-chip data reuse. However, the auxiliary data is too large to be fully retained on-chip, resulting in an off-chip memory bottleneck when applied to applications with large multiplication depth. 
More recent researches~\cite{Trinity,alchemist,UFC}, motivated by the low utilization of CKKS computational modules when applied to TFHE ciphertexts, have enhanced the configurability of compute units based on existing CKKS architectures, achieving higher computational efficiency for both schemes. Still, the performance of these accelerators in CKKS applications remains constrained by off-chip memory access. 

\begin{figure}[]
    \centering
    \includegraphics[scale=0.28]{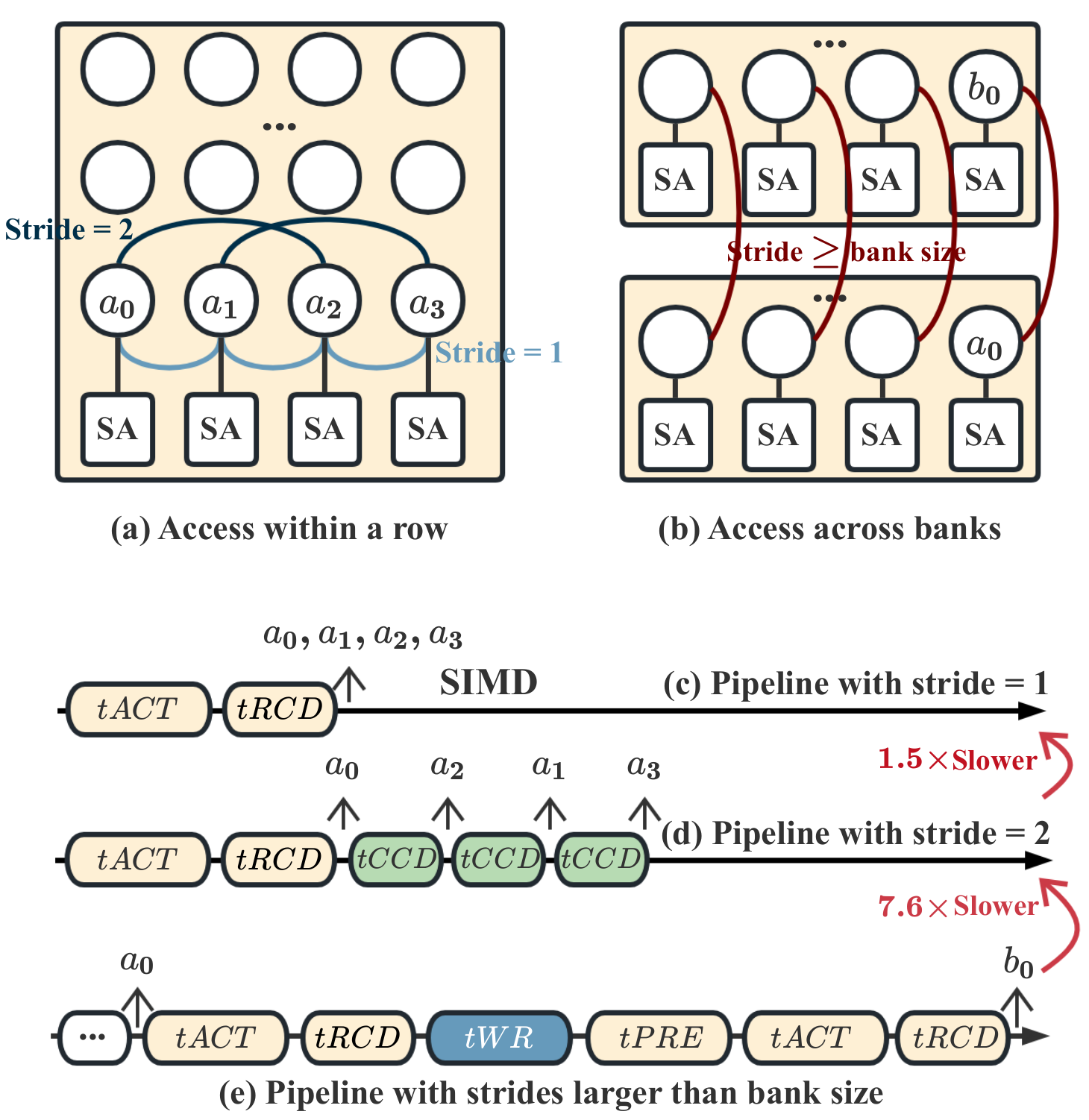}
    \caption{FHE exhibits varying strides of memory access within subarray rows (a) or across banks (b). Different memory access strides result in various pipelines and timing efficiency. A case of DRAM timing: tACT (row activating, 12 ns), tWR (row writing back, 8 ns), tRCD (row to column delay, 12 ns), tCCD (column to column delay, 4 ns), tPRE (row precharging, 12 ns). }
    \label{fig:memory access strides}
    \vspace{-0.5cm}
\end{figure}

DRAM-based Near-Memory Processing (NMP) emerges as a promising candidate for FHE due to the extensive in-memory bandwidth and large memory capacity. 
However, existing NMP FHE accelerators struggle to fully utilize the substantial in-memory bandwidth. For example, APACHE~\cite{APACHE} proposes to deploy its main computational units at the rank level, leveraging parallel ranks to enhance off-chip bandwidth. Despite this, APACHE only achieves an off-chip bandwidth similar to the HBM bandwidth of existing ASIC accelerators, remaining a critical bottleneck in CKKS application. 
Considering the substantial in-memory bandwidth available at the subarray level within DRAM chips, FHEmem~\cite{FHEmem} integrates PEs near each subarray and facilitates communication through a hierarchical interconnect bus design. Nevertheless, its near-subarray design fails to efficiently accommodate the varying memory access strides of FHE operators, leading to underutilization of subarray bandwidth. Specifically, in logistic regression training~\cite{HELR}, read/write latency accounts for 21\% of the total execution time, and the overhead of in-memory data movement through the topologies constitutes 44.7\%. As a result, the architecture struggles to effectively leverage in-memory bandwidth and suffers from significant memory access overhead.

Despite DRAM's potential of extensive bandwidth at the subarray level, a key challenge emerges regarding high bandwidth utilization: \textbf{mismatch between the varying memory access strides of FHE and the fixed access patterns of subarrays.} Specifically, DRAM accesses data by activating and storing a single subarray row in sense amplifiers (SAs) through bitlines. The memory access with stride 1 supports preserving the original order within the subarray row, thereby allowing for SIMD accessing, as in Fig.~\ref{fig:memory access strides}(c). 
Unfortunately, FHE computes with varying memory access strides within a single subarray row, as depicted in Fig.~\ref{fig:memory access strides}(a). This necessitates multiple column accesses to retrieve and reorder the required data. Such column access intervals, as demonstrated in the example in Fig.~\ref{fig:memory access strides}(d), lead to a $1.5\times$ degradation of memory performance. 
Furthermore, FHE also exhibits a large memory access footprint which spans multiple banks or chips shown in Fig.~\ref{fig:memory access strides}(b), leading to an expensive in-DRAM data movement. As illustrated in Fig.~\ref{fig:memory access strides}(e), cross-bank data transfer incurs substantial memory access latency as row activating and writing back when facilitated through inter-bank buses~\cite{RowClone}. 
Therefore, despite the extensive bandwidth provided by subarrays, FHE remains memory-bounded, similar to that in traditional architectures.

To address the issues, we propose FlexMem, a high-parallel NMP architecture that supports flexible dataflow in FHE. 
\textbf{First}, we design a homogeneous architecture that supports fundamental FHE operators, featuring flexible memory access patterns, which locally support varying memory access strides. 
\textbf{Second}, we support the memory access strides across an entire subarray row, different banks, and higher levels of DRAM hierarchy through hierarchical interconnection designs. 
\textbf{Third}, we propose a flexible data management method that dynamically adjusts the layout based on the polynomial-level parallelism at runtime, achieving high hardware utilization. Furthermore, based on the architecture, we design dataflows at the ciphertext level, fully exploiting the parallelism available in both hardware and algorithms.

Our contributions are summarized as follows:\begin{itemize}
\item We design a high-parallel homogeneous architecture that provides unified support for all fundamental operators of both FHE schemes. Leveraging hierarchical topology designs, we efficiently handle the varying memory access strides and make full use of near-memory bandwidth. 
\item We propose a dataflow management method at FHE polynomial- and ciphertext-level, which achieves high hardware utilization and parallel performance. 
\item We conduct a comprehensive evaluation of FlexMem on CKKS, TFHE, and hybrid scheme benchmarks. Compared to the state-of-the-art traditional and near-memory accelerators, Trinity~\cite{Trinity} and APACHE~\cite{APACHE}, our work demonstrates an average speedup of $1.51\times$ and $1.12\times$, respectively. 
\end{itemize}

\section{Background}\label{sec:background}
In this section, we first summarize and analyze the computational pattern of the polynomial-level operators in CKKS and TFHE schemes, and ciphertext-level applications in CKKS. We then introduce the background on DRAM basic operations and in-DRAM data movement techniques. 

\subsection{Polynomial-level operators}\label{sec:polynomial-level operators}

\noindent \textbf{NTT and INTT.}
Number-theoretic transform (NTT) and its inverse transformation (INTT) are both important operators for efficient polynomial multiplication. Both NTT and INTT follow a stage-by-stage computational process in the previous implementations~\cite{scalableNTT, SHARP, ARK}. At each stage, butterfly operations are applied to all pairs of coefficients, following the \eqref{eq NTT}. 
Besides, the distance $d$ between the two coefficients in each pair varies with the stage index $s$ according to $d = 2^s$ for NTT.

\begin{equation}\label{eq NTT}
\begin{aligned}
a_i^{s+1} &=(a_i^s + a_{i+d}^s \cdot \omega)\bmod q, \\
a_{i+d}^{s+1} &=(a_i^s - a_{i+d}^s \cdot \omega)\bmod q  \\
\end{aligned}
\end{equation}
$a_i^s$, $a_{i+d}^s$ are coefficients in stage $s$, and $a_i^{s+1}$, $a_{i+d}^{s+1}$ are coefficients in stage $s+1$. $\omega$ is a pre-computed input. 

\noindent \textbf{BConv.}
Basis Conversion (BConv) converts a ciphertext $x$ from modulus $Q$ to $P$ in two steps detailed in \eqref{eq BConv second step}. BConv primarily performs multiple parallel reductions on vectors. 

\begin{equation}\label{eq BConv second step}
    \begin{aligned}
     tmp_i = \left[\left[x\right]_{q_i} \cdot \hat{q}_i^{-1}\right]_{q_i}, \;
     \left[x\right]_{p_j} = \sum_{i=0}^{l-1} \left[tmp_i \cdot \hat{q_i}\right]_{p_j}, \, 0 \leq j <k \\
    \end{aligned}
\end{equation}
$l$ and $k$ are the number of modulis in $Q$ and $P$, respectively.

\subsection{Ciphertext-level operations in CKKS}\label{sec:ciphertext-level operations in CKKS}
We introduce the fundamental computation pattern of CKKS at the ciphertext level. To this end, we present operations commonly used in CKKS applications~\cite{Bootstrapping, HEResNet, LoLa}. 

\begin{figure}[]
    \centering
    \includegraphics[scale=0.18]{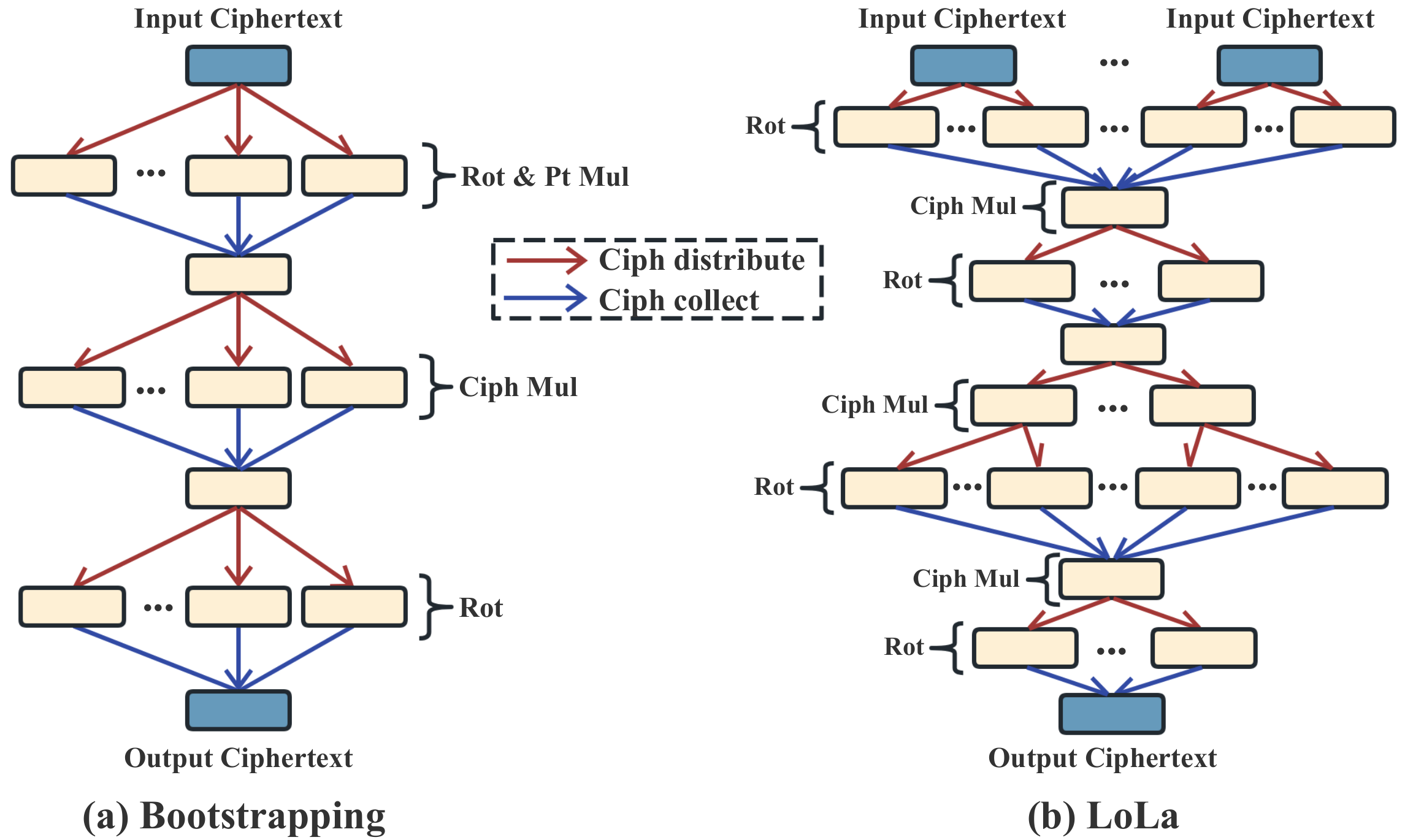}
    \caption{Ciphertext-level computation process in CKKS. }
    \label{fig:ciphertext-level computation process}
    \vspace{-0.3cm}
\end{figure}

\noindent \textbf{Bootstrapping.} This is a critical operation in CKKS~\cite{Bootstrapping}, refreshing a ciphertext that has exhausted its multiplicative depth. The main computations consist of three steps in Fig.~\ref{fig:ciphertext-level computation process}(a). Each step primarily involves parallel ciphertext rotations or multiplications. 

\noindent \textbf{Others.} CKKS-based applications, such as LoLa~\cite{LoLa} in Fig.~\ref{fig:ciphertext-level computation process}(b), exhibit similar ciphertext-level computation patterns. Within each computation stage, multiple ciphertexts perform identical computations such as rotations or multiplications without data dependencies. Before advancing to the next stage, ciphertexts are duplicated or reduced, exhibiting inter-stage data dependency.

\subsection{DRAM operations}\label{sec:DRAM operations}
\begin{figure}[]
    \centering
    \includegraphics[scale=0.20]{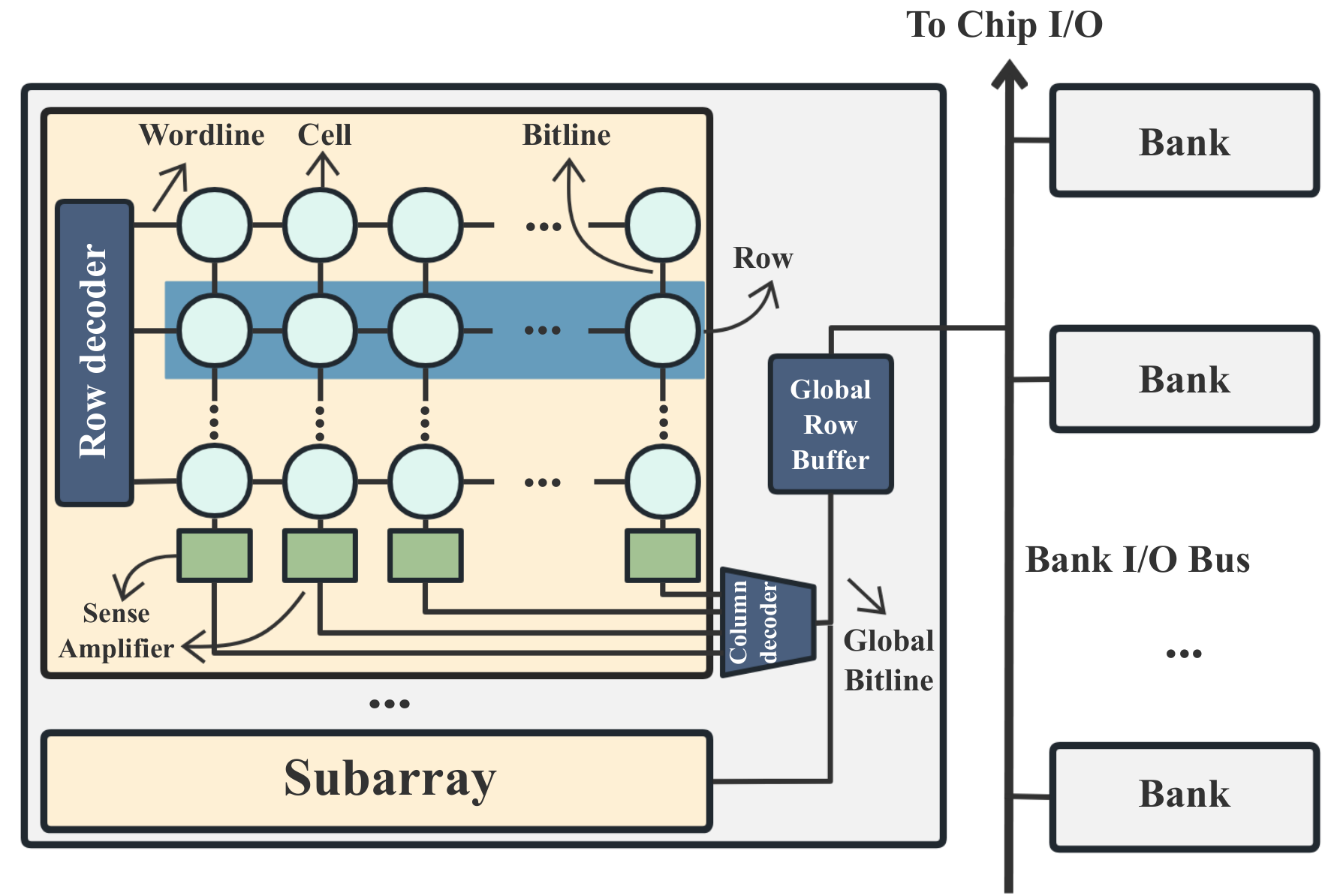}
    \caption{Logical organization inside a chip. }
    \label{fig:subarray}
    \vspace{-0.3cm}
\end{figure}

The high-level memory hierarchy of DRAM consists of channels, DIMMs, and chips. The organization of each chip is illustrated in Fig.~\ref{fig:subarray}, comprising multiple banks, with each bank consisting of several subarrays. Data are stored in cells in subarrays. Accessing data in a subarray involves the following steps. First, the row decoder activates a row based on the given address. After a latency of $tACT + tRCD$, the row is maintained in SAs. Subsequently, the global row buffer (GRB) retrieves contiguous data selected by the column decoder after the $READ$ command is triggered. Multiple $READ$ commands targeting the same row can sequentially access data from SAs, with an interval of $tCCD$. After the row access completes or when accessing a different row, the row must be precharged. Precharging requires a latency of $tPRE$ until the next row activation. 

\subsection{In-DRAM data movement}\label{sec:In-DRAM data movement}
Existing works have explored efficient in-DRAM data movement at different hierarchies. For example, LISA~\cite{LISA} introduces isolation transistors between adjacent subarrays to enable high-parallel data transfers across adjacent subarrays. Figaro~\cite{figaro} enables fine-grained data movement between subarrays based on the observation that all subarrays within a bank share the GRB through the global bitline. RowClone~\cite{RowClone} utilizes inter-bank data buses to facilitate serial data transfers across banks. At a high level of the DRAM hierarchy, DIMM-link~\cite{DIMM-link} proposes a routing and interconnection mechanism to enable data movement between DIMMs. In our work, we adopt the designs of LISA and DIMM-link to accommodate subarray and DIMM-level in-DRAM data movement.

\section{Motivation}\label{sec:motivation}

\subsection{Why near-memory processing for FHE}
Many studies~\cite{FAB,craterlake,ARK,SHARP} have highlighted that the substantial memory access demands of CKKS lead to a critical bottleneck. Despite algorithmic optimizations~\cite{MAD,ARK} and the use of large scratchpad memory, the improvement remains limited. Inefficiency is exhibited regarding the off-chip bandwidth and on-chip capacity utilization. In bootstrapping~\cite{Bootstrapping}, which is a memory-bounded primitive, CraterLake~\cite{craterlake} shows that the effective utilization of off-chip bandwidth remains low, with more than half being consumed by intermediate results evicting and reloading.  
Besides, Sharp~\cite{SHARP} demonstrates that although on-chip memory plays a significant role in acceleration, occupying approximately 45.1\% of the chip area, its utilization throughout the application remains low (35\%). 

The issue of scratchpad and off-chip bandwidth utilization essentially arises because of the extensive and frequently used auxiliary data. To accommodate auxiliary data off-chip loading, intermediate results have to be evicted and reloaded in subsequent computations. Therefore, only a limited portion of off-chip bandwidth is effectively allocated to ciphertext operands. 
Besides, the expensive bootstrapping, which involves a large amount of auxiliary data, does not occur uniformly throughout the entire process but rather periodically. Therefore, existing accelerators have to adopt large on-chip memory for high bootstrapping performance. However, the utilization of on-chip memory remains relatively low during most of the computation process. 
The limitations of the architecture motivate us to explore a DRAM-based near-memory architecture, which offers sufficient capacity to store all auxiliary and intermediate data while providing substantial near-memory bandwidth, presenting a promising solution for alleviating the memory bottleneck.

\subsection{Challenge of supporting FHE in DRAM}\label{sec:challenge} 
\noindent \textbf{Challenge in integrating NMP units.}
NMP units in DRAM can be integrated at different hierarchical levels. 
Existing works have adopted high-level DRAM hierarchy integrating, such as near-DIMM~\cite{DIMM_NMP} and near-rank designs~\cite{APACHE}. Although these designs achieve high acceleration performance with minimal modifications to the DRAM module, the limited near-memory bandwidth necessitates the integration of large register files (RFs) within accelerators to enhance data reuse. The memory hierarchy of these works is similar to conventional accelerators, nevertheless, their RFs feature larger area budgets, occupying 71.6\% of the total area~\cite{APACHE}, which results in a relatively lower area budget for computational units, making it challenging to achieve high efficiency.

\noindent \textbf{Challenge of supporting various memory access strides.}
Recent studies have integrated computational units into lower levels of memory hierarchy, such as near-subarray designs~\cite{FHEmem,NTT-PIM}, which can provide sufficient bandwidth to support the parallelism required by FHE and reduce the reliance on extra registers. However, the regular memory access pattern of subarrays makes it difficult to fully utilize near-memory bandwidth when handling FHE operators with varying memory access strides. On the one hand, a varying stride within a single subarray row leads to additional column access intervals, as multiple $READ$ commands are required to access the activated row in SAs at the granularity of the GRB size. The issuing of each $READ$ command must wait for $tCCD$ interval, as defined in the JEDEC standard~\cite{jedec}. 
On the other hand, when the memory access stride exceeds the size of a bank or a chip, computational units experience long idle periods while waiting for time-consumed intra-memory data transfers to retrieve the required operands. Specifically, data movement at the bank level can be achieved through the bank I/O bus. Such movement is performed serially and is constrained by the bit-width of GRB size and DRAM timings. Besides, chip-level data transfer does not have a direct datapath implementation in DRAM and is typically accomplished through the CPU. Due to the limited DRAM I/O bandwidth, this incurs significant latency. 
Therefore, the varying memory access strides of FHE exacerbate memory access inefficiencies, thereby hindering the effective utilization of near-memory bandwidth.

\noindent \textbf{Challenge of supporting flexible dataflow.}
The flexible dataflows in FHE encompass multiple levels. Basic operators compute at the polynomial level. For instance, modular element-wise operators and NTT exhibit parallel computations at the polynomial level, and BConv performs reduction across polynomials. Therefore the inherently varying number of polynomials in CKKS complicates the dataflow and poses a challenge for operators to fully utilize the near-memory bandwidth. 
Furthermore, CKKS demonstrates ciphertext-level parallelism. However, as illustrated in Fig.~\ref{fig:ciphertext-level computation process}, ciphertexts exhibit inherent data dependencies throughout the computation. Given their substantial size (e.g., 13.5 MB each ciphertext), the movement of ciphertexts across memory incurs significant overhead. Consequently, effectively harnessing ciphertext-level parallelism presents a key challenge, particularly in balancing computation with the cost of ciphertext communication. 

\section{FlexMem architecture}\label{sec:FlexMem architecture}
\subsection{Architecture overview}\label{sec:architecture overview}

\begin{figure}[]
    \centering
    \includegraphics[scale=0.30]{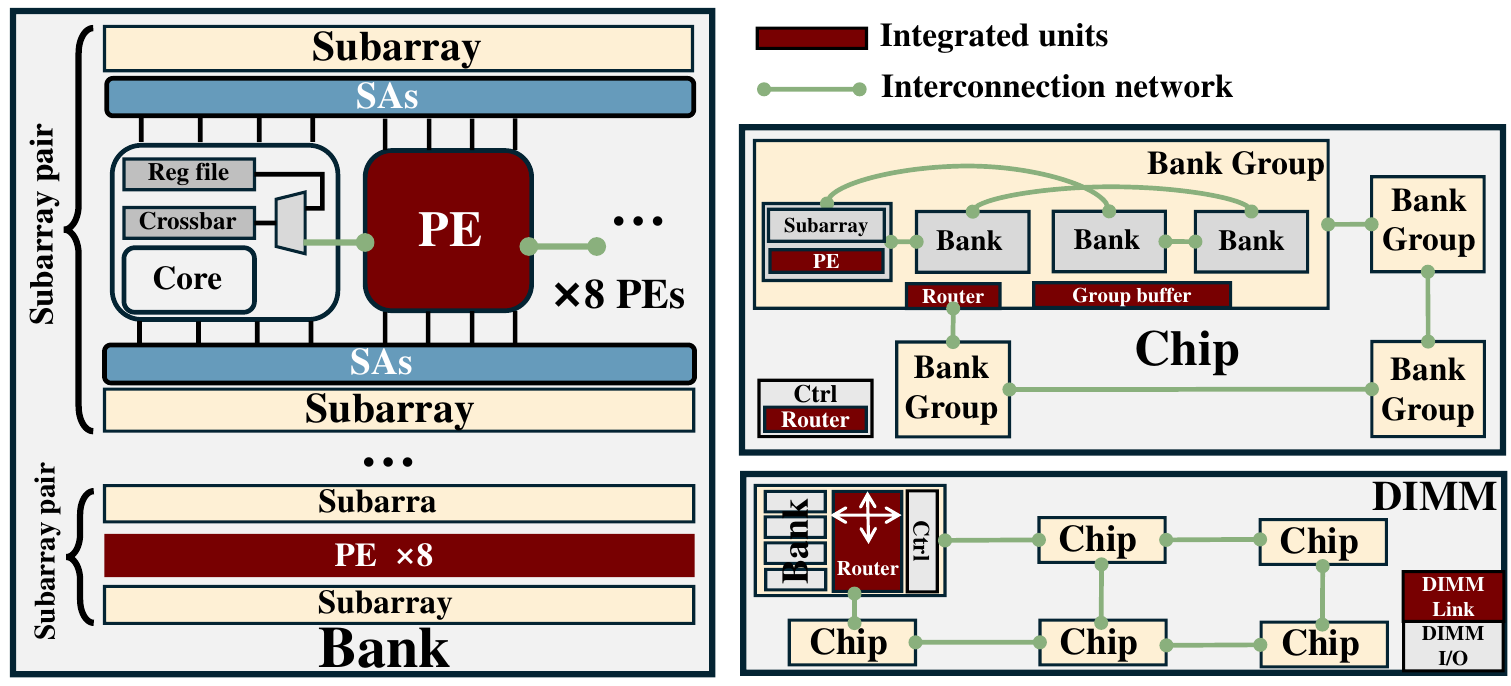}
    \caption{Overall architecture. PEs are integrated between each subarray pair. Parallel pathways among banks are constructed for NTT inter-stage coefficient switching. Chip-level network is proposed for ciphertext-level data transfer. }
    \label{fig:overview}
    \vspace{-0.2cm}
\end{figure}

The overall architecture is depicted in Fig.~\ref{fig:overview}. FlexMem exploits the parallelism of DIMM, chip, bank, and subarray levels in the DRAM hierarchy. Our near-subarray PE is designed within each bank inside a chip and can be scaled to multiple chips and DIMMs to accommodate various applications. The entire design primarily consists of the following components:\begin{itemize}
    \item \textbf{Intra-bank PE.} Multiple homogeneous processing elements (PEs) are integrated between two adjacent subarrays within a bank. Each PE includes a computing core, RFs, a crossbar, and the chaining of adjacent PEs. Within the core, the fundamental operators of FHE are implemented based on modular multiplication, addition, and subtraction. 
    \item \textbf{Inter-bank network.} We group banks and design a leap-based network along with corresponding routers both within and across groups to enable efficient inter-bank data movement. 
    \item \textbf{Inter-chip network.} We propose a lightweight network routing design based on memory access patterns of ciphertexts. To further enhance scalability, we implement a DIMM-link~\cite{DIMM-link} mechanism between adjacent DIMMs, enabling FlexMem for larger-scale applications. 
\end{itemize} 

\subsection{PE design}\label{sec:PE design}
\begin{figure}[]
    \centering
    \includegraphics[scale=0.26]{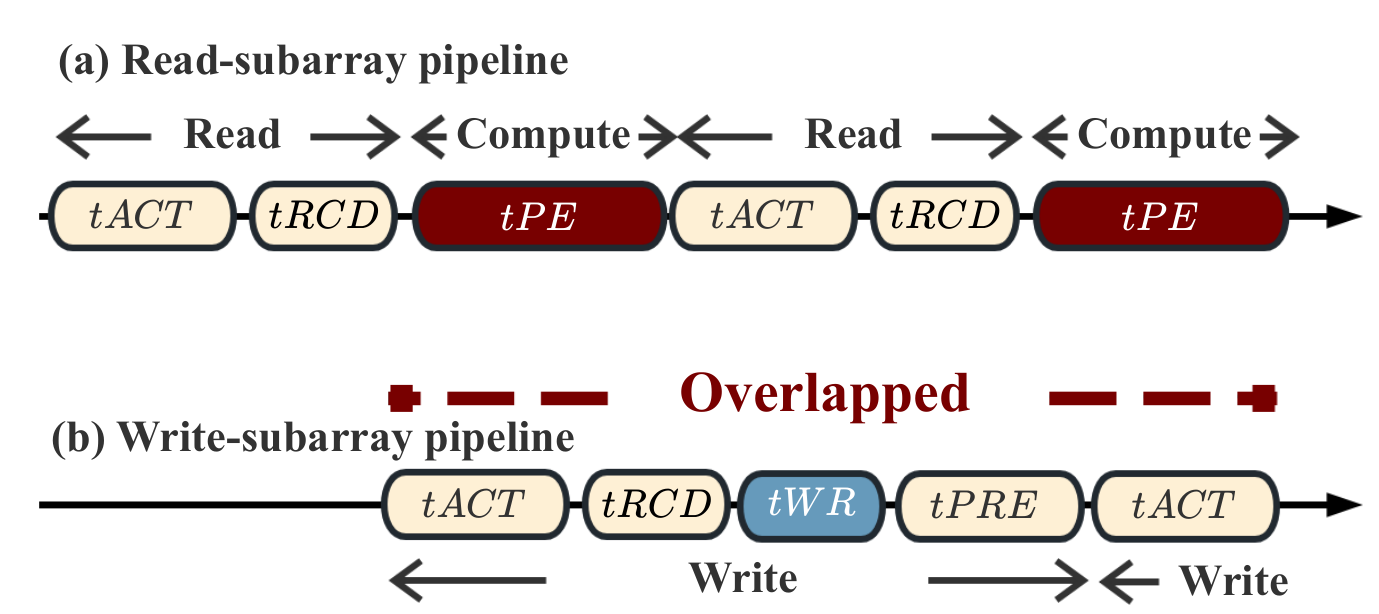}
    \caption{Pipeline between adjacent subarrays. }
    \label{fig:subarray pipeline}
    \vspace{-0.5cm}
\end{figure}
Subarray accesses typically incur long latency. Specifically, reading and writing to the same subarray inherently cannot occur simultaneously, with tens of cycles latency of precharging and writing back in between. Therefore, we adopt a ping-pong buffer organization with two adjacent subarrays. PEs are deployed between the subarray pairs, which allows us to continuously read from one subarray while writing to the other, thereby overlapping the read and write latency, as shown in Fig.~\ref{fig:subarray pipeline}. 
Moreover, varying memory access strides within a single subarray row result in a large portion of column access intervals. To address the issue, we build on a key insight that, \textbf{the optimal row access performance is attained only when the internal data order remains unchanged.} We reformulate the challenge of variable access strides within a single subarray row as a combination of intra-PE access and inter-PE communication. 

\subsubsection{Intra-PE access} 
Each PE is equipped with a 64-to-64 crossbar and 0.5 KB RFs, which are directly connected to SAs, allowing us to preserve the original row order while enabling flexible intra-PE memory access. 

\subsubsection{Inter-PE communication}
To support memory access strides spanning an entire subarray row, which may occur in BConv or NTT, we design a 32 bit/cycle datapath between adjacent PEs. Each PE first reads data from the subarray, performs local computations, and then exchanges data with other PEs through PE chains for further computations. Throughout the process, the PE core accesses data from either subarray or other PEs. All PEs in a chip operate in a synchronized manner, executing identical computations or communications under a unified control logic. 

\begin{figure}[]
    \centering
    \includegraphics[scale=0.19]{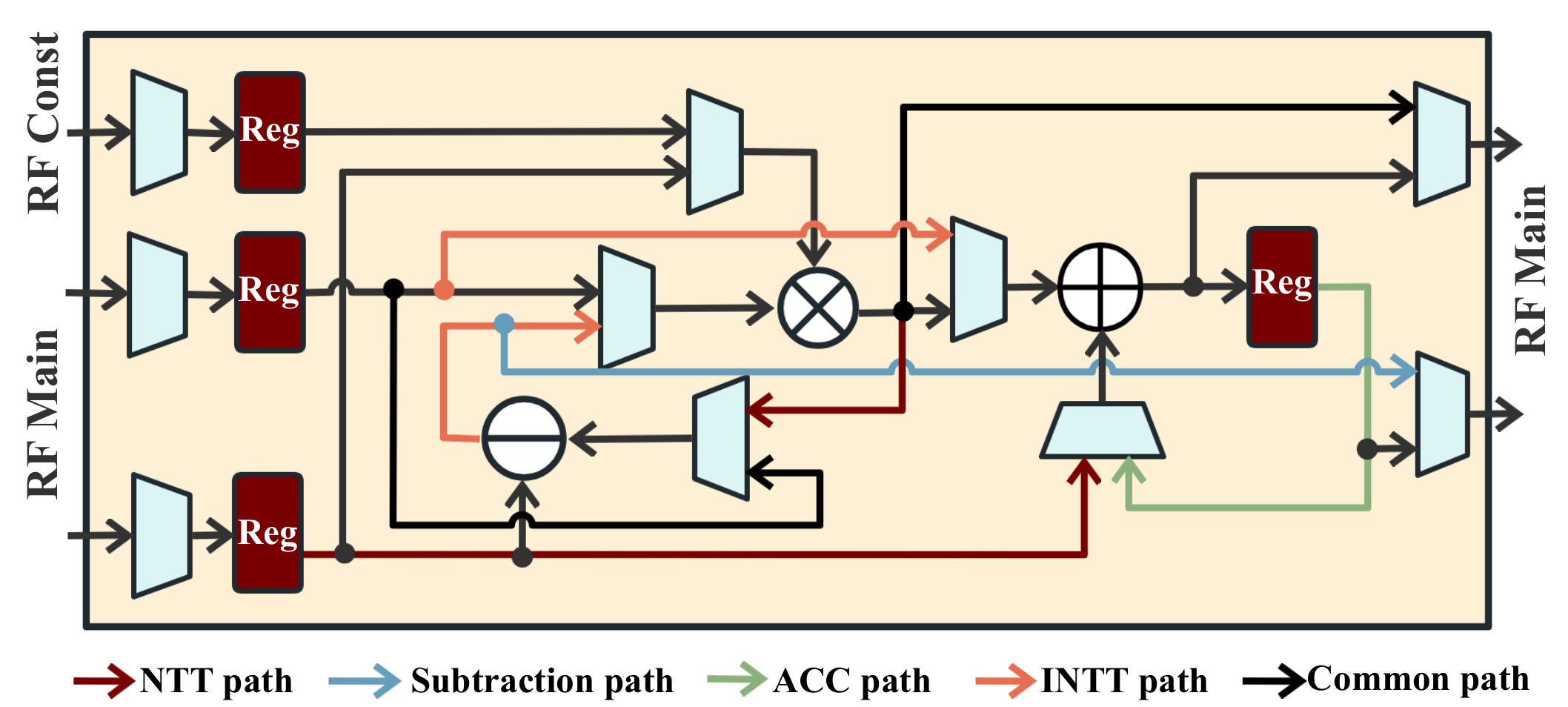}
    \caption{Detailed architecture of our PE core. Computational modules in the core are based on modular arithmetic.}
    \label{fig:PE Core}
    \vspace{-0.3cm}
\end{figure}

Our PE core can support all operators, with architecture depicted in Fig.~\ref{fig:PE Core}. FlexMem is a homogeneous and distributed design with multiple PEs. FHE basic operators follow a similar dataflow paradigm between adjacent subarrays, as long as the required operands are entirely within the subarray. However, FHE involves memory access across multiple subarrays, banks, and even chips, necessitating data movement before feeding the aforementioned computation pipeline. To address inter-subarray transfer, we draw inspiration from LISA's design~\cite{LISA}. We further elaborate on our method in the following sections for bank and chip-level communication. 

\subsection{In-memory communication design}\label{sec:in-memory communication}
Under our data layout, bank-to-bank communication is essential for NTT, and chip-to-chip communication is required for higher-level operations such as bootstrapping. 
However, existing approaches utilize the shared bus between banks~\cite{RowClone}, facilitating serial communication among banks. This method is inefficient for NTT, where every two banks need to communicate simultaneously. Moreover, current chip-to-chip communication in DRAM relies on the host CPU for transferring, which is limited by external bandwidth, making in-memory data movement a critical bottleneck. To address these inefficiencies, we propose bank and chip-level communication mechanisms driven by FHE memory access patterns. We will further elaborate in Sec~\ref{sec:Dataflow management} on how the topology designs support the specific dataflow.

\subsubsection{Bank-level}
The NTT memory access stride increases exponentially with each stage, resulting in a complex communication pattern among different banks. Therefore, simply adopting the the partial-chain approach in~\cite{FHEmem} would require multiple routing steps, leading to a long latency. 

To tackle this issue, we design highly parallel pathways for inter-bank dataflow of coefficient switching, which directly links pairs of banks according to the NTT strides in the first few stages. 
Instead of connecting all banks within the memory access stride, we adopt a hardware-efficient method by grouping banks. Within each group, banks are skip-connected, and each group is also linked with others using the same skip-connection, as illustrated in Fig.~\ref{fig:overview}. This design ensures that data transfers between closely located banks can be completed in a single routing step, while transfers between distant banks require at most two routing steps.

\subsubsection{Chip-level}
CKKS applications exhibit similar computational patterns at the ciphertext level, as described in Sec~\ref{sec:ciphertext-level operations in CKKS}, where each stage exhibits parallel ciphertext operations. To fully leverage the inherent parallelism, we distribute parallel ciphertexts across different chip pairs. 
Because of the ciphertext-level inter-stage data dependencies, the data movement among chips incurs extensive overhead when accomplished through the host CPU, accounting for 41.26\% of the latency in FlexMem without chip-level interconnection. 
Based on an observation that, \textbf{the ciphertext-level movement patterns primarily consist of distribution and collection}. We employ a mesh topology of linking chips, as depicted in Fig.~\ref{fig:overview}. Our design enables simultaneous inter-chip transfer and intra-chip computation, overlapping the transfer latency and reducing the chip-level transfer portion to 8.00\%. 
For ciphertext transfer between DIMMs, we adopt DIMM-link~\cite{DIMM-link} between adjacent DIMMs to extend our work to a larger scale. 

\section{Dataflow management}\label{sec:Dataflow management}
\subsection{Data layout}\label{sec:data layout}
\begin{figure}[]
    \centering
    \includegraphics[scale=0.36]{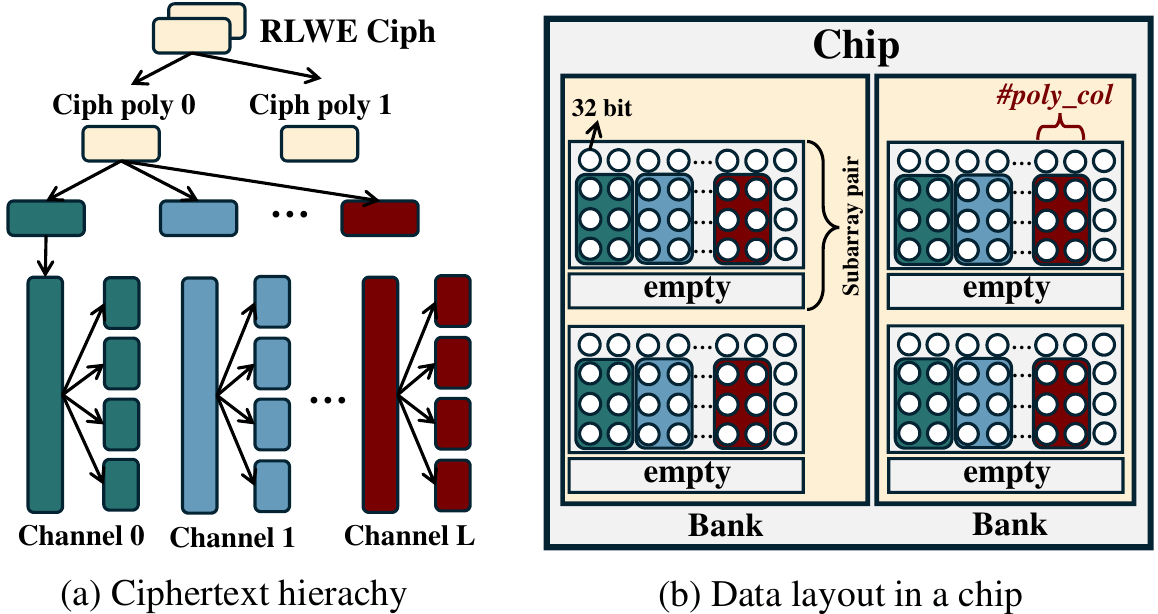}
    \caption{A toy example of a data layout of an RLWE ciphertext with L + 1 RNS channels. }
    \label{fig:data layout}
    \vspace{-0.2cm}
\end{figure}

Our core data layout strategy is designed to fully exploit parallelism. An RLWE ciphertext, which consists of two polynomials with modulus $Q$, is allocated to a chip pair. Within a chip, each ciphertext polynomial, composed of multiple RNS channels, is evenly tiled along the coefficient dimension and distributed across all banks. Within each bank, the tiles are further split across half the subarrays. Finally, tiles in each subarray are arranged into multiple columns. The number of 32-bit columns occupied by each tile is denoted as $\#poly\_col$. This is a runtime-configurable parameter in our design, which is set to ensure that the tiles of all RNS channels can be accommodated within a single subarray. An example is shown in Fig.~\ref{fig:data layout}. 
For each LWE ciphertext, we organize its layout as an RNS channel of an RLWE ciphertext and make sure that all data required in a programmable bootstrapping (PBS) is within a chip. 
Besides, we arrange twiddle factors and bootstrapping plaintexts in the same chips with the ciphertexts they are computed with. For evaluation keys used for ciphertext rotations and multiplications, we redundantly stored across multiple chips because of the parallel execution of ciphertext-level operations. 

\subsection{Polynomial-level dataflow design}\label{sec:dataflow mapping}
We support basic operators by leveraging PEs and intra-chip networks. To further illustrate the polynomial-level dataflow in FlexMem, we take NTT and reduction as examples. 

\begin{figure}[]
    \centering
    \includegraphics[scale=0.27]{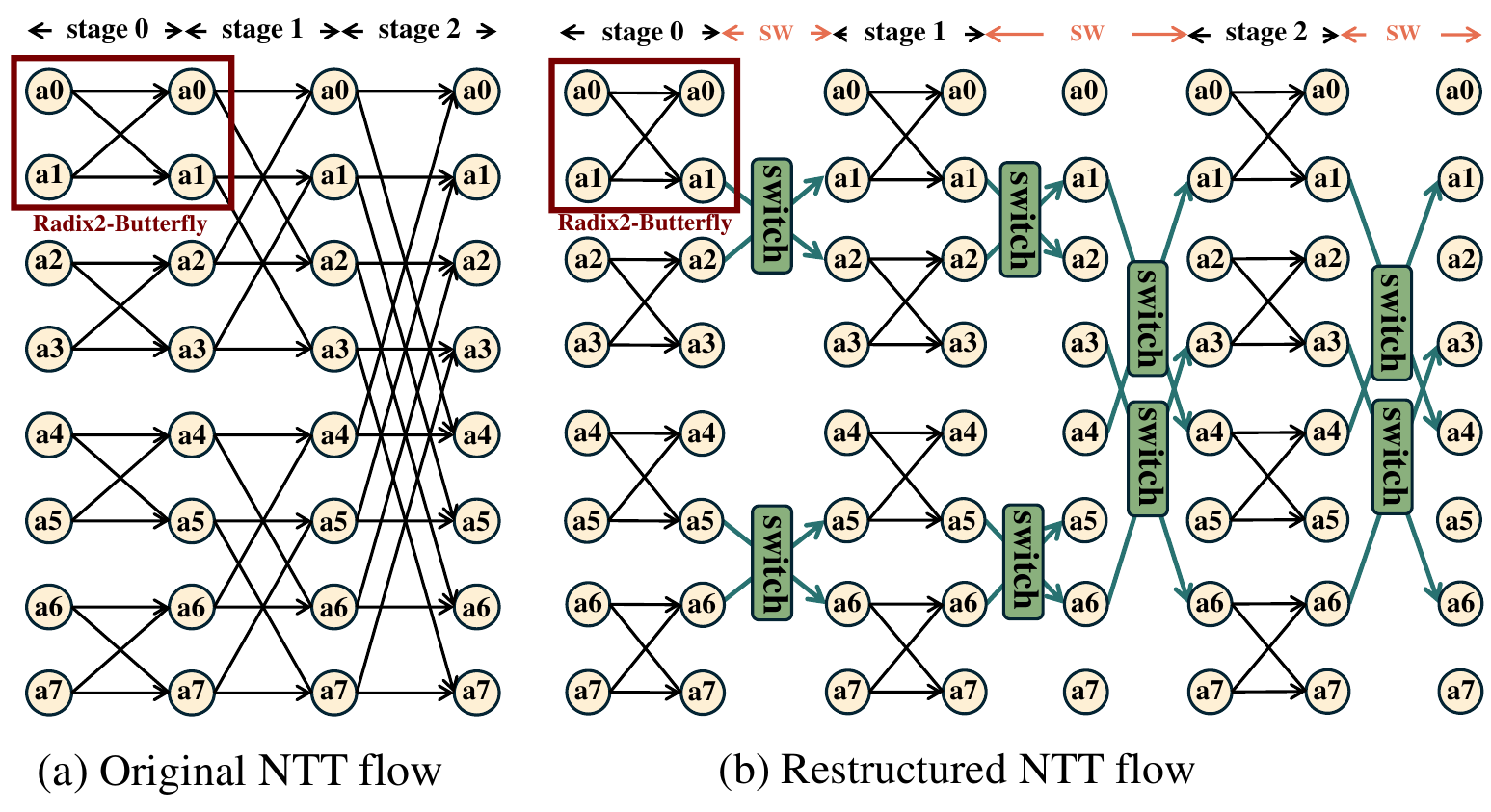}
    \caption{An example of NTT with coefficients switched after each stage to maintain the same radix-2 butterfly computational pattern in each stage. }
    \label{fig:NTT flow}
    \vspace{-0.5cm}
\end{figure}

\noindent \textbf{NTT.}
We note that although the 4-step NTT in existing works~\cite{FAB,ARK,craterlake,SHARP,Trinity} improve the throughput via pipelining optimization, it introduces additional computation, with $12.5\%$ more computation in $2^{16}$-points NTT and $16.7\%$ for $2^{12}$-points NTT because of the additional modular multiplication between the two sub-NTTs. Therefore, we adopt a unified radix-2 NTT with a fixed memory access pattern in each stage through coefficient switching, as depicted in Fig.~\ref{fig:NTT flow}. This approach preserves the same computational complexity as the original NTT algorithm. We perform coefficient switching after each stage, where the distance between the switched coefficients increases exponentially with the stage index $s$, following $d = 2^s$. We accomplish the coefficient switching through hierarchical topologies in FlexMem. Taking a $2^{16}$-points NTT with 16 stages as an example. \textbf{Step 1:} Each PE locally reads 32 coefficients and internally completes the first 5 stages. \textbf{Step 2:} The 8 PEs of each subarray pair exchange coefficients through the PE chain and complete the next 3 stages. \textbf{Step 3:} The 16 subarray pairs within each bank communicate with other subarrays through LISA link~\cite{LISA} to conduct the next 4 stages. \textbf{Step 4:} Finally, to finish the last 4 stages, 16 banks in each bank group communicate with each other after each stage of computation through the bank-level interconnection. 

\noindent \textbf{Reduction.}
Reduction is the core computation in BConv and PBS, accounting for more than $95\%$ of the total computation. In FlexMem, the overall reduction resembles an adder tree process. Each PE follows the ACC path in Fig.~\ref{fig:PE Core}, to complete the local reduction which corresponds to the computation at the leaf nodes of the adder tree. Subsequently, PE forwards partial results to other PEs within subarray pairs, where each PE corresponds to a node in the adder tree.
The reduction proceeds iteratively until all coefficients are reduced, with coefficients being transferred through the PE-chain. Since all RNS channels in RLWE and all LWE ciphertexts in TFHE PBS are allocated within the same subarray, the reduction does not require additional inter-subarray or inter-bank data exchange.

\subsection{Polynomial-level dataflow tuning}\label{sec:polynomial-level dataflow tuning}
We propose a data remapping method based on the key observation that FlexMem suffers from performance degradation because of the frequent changes in polynomial-level parallelism, leading to low near-memory bandwidth utilization. 
Remapping can change the number of columns for polynomial tiles in a subarray, i.e., $\#poly\_col$. For operators that change the number of polynomials, such as BConv, we remap either before the operator, reducing $\#poly\_col$ to allocate more columns for subsequently added polynomials, or after the operator that discards polynomials, thereby increasing $\#poly\_col$ to fill more coefficients in a subarray row. With remapping integrated into the computation flow, we adaptively adjust the polynomial-level dataflow, maintaining near-full utilization of the bandwidth. 

\begin{figure}[]
    \centering
    \includegraphics[scale=0.43]{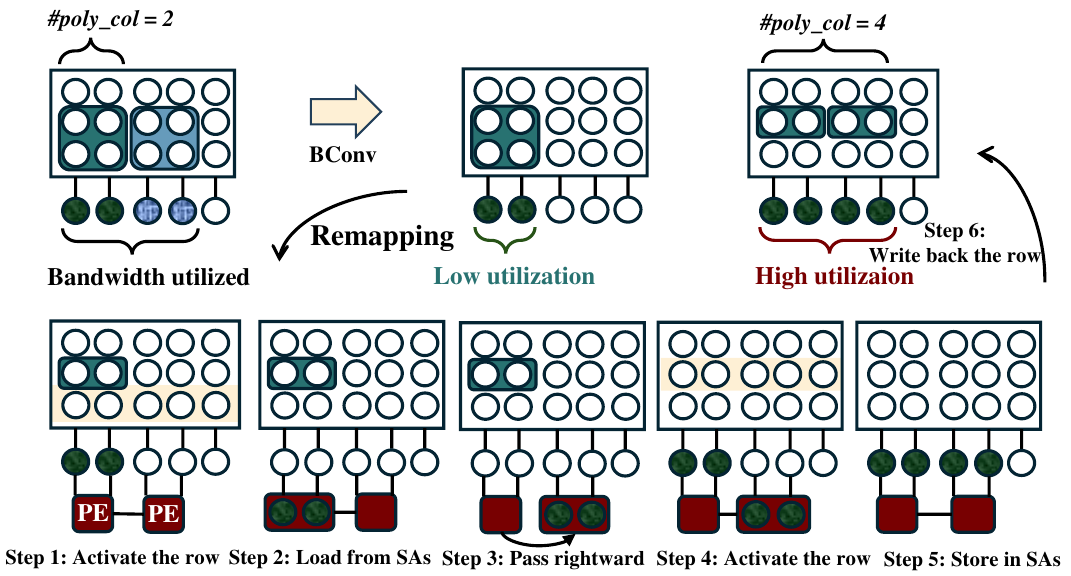}
    \vspace{-0.1cm}
    \caption{A toy example of remapping to maintain high bandwidth utilization after discarding several channels in BConv. }
    \label{fig:remapping}
    \vspace{-0.3cm}
\end{figure}

Remapping is conducted primarily through PE chains. Each remapping either doubles or halves the $\#poly\_col$. For example, as illustrated in Fig.~\ref{fig:remapping}, to expand the number of columns, the lower half of rows in a subarray are activated, starting from the bottom row in a pipelining manner. The data is then read into computational units, shifted rightward, and written back to the subarray. We show that the remapping is highly efficient in CKKS applications, which accounts for merely $0.68\%$ of the latency, significantly boosting bandwidth utilization and leading to $2.00\times$ of performance improvement in bootstrapping. 

\subsection{Ciphertext-level dataflow design}\label{sec:ciphertext-level dataflow design}
We further elaborate on the dataflow of high-level FHE operations. For TFHE, PBS for multiple ciphertexts can either be executed serially within a single chip to enhance data reuse or distributed across multiple chips in parallel for high throughput. 

For CKKS, the core operator key switching is inherent parallel across two ciphertext polynomials and, therefore, allocated to a pair of chips. Higher-level applications that are built on key switching, are distributed across multiple chip pairs. 
We allocate an appropriate number of chips based on the ciphertext-level parallelism for a given task. 
For example, in bootstrapping with ARK~\cite{ARK} parameters, each baby-step computation involves eight parallel ciphertext rotations, which are allocated into four chip pairs, each handling two ciphertext rotations. 
Ciphertext-level communication patterns primarily consist of distribution and collection. The distribution phase begins with the first chip pair, where ciphertext-level computation is performed while simultaneously forwarding the input ciphertext to the next chip pair. Each chip pair starts its computation upon receiving the input ciphertext while forwarding it to the next pair. This continues until the last chip pair completes its task. In the collection phase, each pair first performs intra-chip reduction, then iteratively reduces results with its neighboring pair until the final reduction is completed.

\begin{table}[]
    \centering
    \caption{Configuration of DRAM in our design}
    \label{tab:config}
    \fontsize{8}{9}\selectfont
    \resizebox{0.45\textwidth}{!}{
    \begin{tabular}{c}
        \toprule
        \textbf{DRAM architecture} \\ \hline
        DDR5, 1 channel, 2 DIMMs per channel, 4 chips per DIMM, 64 banks per chip, \\
        32 subarrays per bank, 1024 row, 2 KB per row buffer, 32 B global row buffer \\
        \bottomrule
        \textbf{DRAM timings (cycles)} \\ \hline
        tCCDS = 4, tCCD = 2, tWR = 8, tPRE = 12, tACT = 24, tRCD = 24, tRRD = 24 \\
        \bottomrule
    \end{tabular}}
\end{table}

\section{Evaluation}\label{sec:evaluation}
\subsection{Hardware configuration}
Our design is based on a DRAM with configurations in Table~\ref{tab:config}. For CKKS, TFHE NN, and hybrid scheme benchmarks, to make a fair comparison, we adopt 2 DIMMs each with 4 DRAM chips. The near-memory designs in this configuration have a similar area to accelerators when scaled to the same technology node. In programmable bootstrapping, we adopt 1 DIMM with 2 chips to compare with TFHE accelerators. In each of our hardware configurations, we integrate 8 PEs in each subarray pair by default. 

\begin{table}[t]
    \centering
    \caption{Area and thermal power of the near-memory design.}
    \label{tab:implement_result}
    \fontsize{5}{5.5}\selectfont
    \resizebox{0.40\textwidth}{!}{
    \begin{tabular}{lrr}
        \toprule
        \textbf{PE Component} & \textbf{Area (mm\textsuperscript{2})} & \textbf{Power (W)} \\ \hline
        Core & 29.27 & 11.56 \\ 
        Register file (4 MB) & 6.40 & 0.49 \\
        Crossbar & 2.41 & 0.14 \\
        PE-chain & 0.03 & 0.10 \\
        Bank-network & 3.28 & 0.99 \\
        \textbf{Total in-chip design} & \textbf{41.39} & \textbf{13.28} \\ \hline
        $4\times$ in-chip design & 165.56 & 53.12 \\
        $4\times$ chip-network & 4.30 & 4.70 \\
        \textbf{Total in-DIMM design} & \textbf{169.86} & \textbf{57.82} \\ \hline
        \textbf{Total (2 DIMMs)} & \textbf{339.72} & \textbf{115.64} \\
        \bottomrule
    \end{tabular}}
\end{table}

\begin{table}[]
    \centering
    \caption{CKKS benchmarks performance (\textnormal{ms}).}
    \label{tab:CKKS_benchmarks}
    \resizebox{0.40\textwidth}{!}{
    \begin{tabular}{lrrr}
        \toprule
        \textbf{Accelerators} & \textbf{Bootstrap} & \textbf{HELR} & \textbf{ResNet-20} \\ \hline
        CPU-baseline~\cite{concrete} & 17.2 s & 356 s & 23 min \\
        Poseidon~\cite{Poseidon} & 127.45 & 72.98 & 2661.23 \\
        TensorFHE~\cite{tensorfhe} & 421.8 & 220 & 4939 \\
        SHARP~\cite{SHARP} & 3.12 & 2.53 & 99 \\
        UFC~\cite{UFC} & 2.64 & 2.11 & 90 \\
        Trinity~\cite{Trinity} & 1.92 & 1.37 & 89 \\
        Alchemist~\cite{alchemist} & 1.68 & 1.22 & \\
        FHEmem~\cite{FHEmem} & 1.36 & 1.27 & 24.8 \\
        APACHE~\cite{APACHE} & 1.12 & 0.8 & \\
        \textbf{FlexMem} & \textbf{1.00} & \textbf{0.64} & \textbf{18.3} \\
        \bottomrule
    \end{tabular}}
    \vspace{-0.5cm}
\end{table}

\subsection{Implementation results}\label{sec:implementation results}
We synthesize and implement near-memory PEs in a 10 nm process technology using the Design Compiler. We use FinCACTI\cite{fincacti} to model DRAM, register files, and topologies wiring. The power of the in-memory interconnection design is set according to GRS~\cite{GRS}. The bandwidth of datapath among PEs, banks, and chips are set as 4 GB/s, 1.6 TB/s, and 2.4 TB/s respectively. Our PE runs at 1 GHz, and the designed components in 2 DIMMs are sized 339.72 mm\textsuperscript{2}, consuming 115.64 W in total, with the default configuration. 
The area and power breakdown are listed in Table~\ref{tab:implement_result}.

\subsection{End-to-end benchmarks}\label{sec:end-to-end benchmarks}
To evaluate the performance, we develop a cycle-accurate simulator and test it with CKKS, TFHE, and hybrid FHE scheme applications. 
\subsubsection{CKKS benchmarks}\label{sec:CKKS benchmarks}
We evaluate the performance of FlexMem on three CKKS applications: bootstrapping~\cite{Bootstrapping}, HELR~\cite{HELR}, and ResNet-20~\cite{HEResNet}. Compared with CPU~\cite{concrete}, GPU~\cite{tensorfhe}, FPGA~\cite{Poseidon}, ASIC~\cite{SHARP,Trinity,UFC,alchemist}, and NMP~\cite{FHEmem,APACHE} designs. The results are shown in Table~\ref{tab:CKKS_benchmarks}. 

\noindent \textbf{Fully-packed bootstrapping~\cite{Bootstrapping}.} We compare the bootstrapping runtime with previous works, under the same bootstrapping parameters as SHARP~\cite{SHARP}. 

\noindent \textbf{HELR~\cite{HELR}.} This is an ML workload training with logistic regression. We use a batch size of 1024 and train for 32 iterations. We perform bootstrapping every three iterations and record the average latency per iteration. 

\noindent \textbf{ResNet-20~\cite{HEResNet}.} This benchmark is a CNN inference based on the CIFAR-10 dataset. The input image is sized 32 $\times$ 32 $\times$ 3. 

\begin{table}[]
    \centering
    \caption{PBS throughput performance (\textnormal{PBS/s}).}
    \label{tab:PBS_benchmarks}
    \resizebox{0.42\textwidth}{!}{
    \begin{tabular}{lrrrr}
        \toprule
        \textbf{Accelerators} & \textbf{SET \uppercase\expandafter{\romannumeral 1}} & \textbf{SET \uppercase\expandafter{\romannumeral 2}} & \textbf{SET \uppercase\expandafter{\romannumeral 3}} & \textbf{SET \uppercase\expandafter{\romannumeral 4}} \\ \hline
        CPU-baseline~\cite{concrete} & 63 & 36 & 12 & \\
        NuFHE~\cite{nufhe} & 2500 & 550 & & \\
        Morphling~\cite{Morphling} & 147615 & 78692 & 41850 &  \\
        UFC~\cite{UFC} & 94182 & 50792 & 22021 & 3346 \\
        Alchemist~\cite{alchemist} & 123000 & 75000 & & \\
        Trinity~\cite{Trinity} & 150015 & 85034 & 45246 & \\
        APACHE~\cite{APACHE} & 500000 & & & \\
        \textbf{FlexMem} & \textbf{493984} & \textbf{217396} & \textbf{103287} & \textbf{16730} \\
        \bottomrule
    \end{tabular}}
\end{table}

\begin{table}[t]
    \centering
    \caption{NN inference performance (\textnormal{ms}).}
    \label{tab:NN_benchmarks}
    \resizebox{0.35\textwidth}{!}{
    \begin{tabular}{lrrr}
        \toprule
        \textbf{Accelerators} & \textbf{NN-20} & \textbf{NN-50} & \textbf{NN-100} \\ \hline
        CPU-baseline~\cite{NN} & 64.60 s & 129.25 s & 263.54 s \\
        Morphling~\cite{Morphling} & 340 & 840 & 1,720  \\
        UFC~\cite{UFC} & 33.26 & 83.53 & 166.67 \\
        Trinity~\cite{Trinity} & 69.86 & 146.26 & 277.13 \\
        \textbf{FlexMem} & \textbf{39.89} & \textbf{96.44} & \textbf{190.7} \\
        \bottomrule
    \end{tabular}}
    \vspace{-0.4cm}
\end{table}

\subsubsection{TFHE benchmarks}\label{sec:TFHE benchmarks}
To evaluate the performance of FlexMem on logical FHE computations, we assess the PBS throughput (PBS/s) and the latency of the NN benchmark. Notably, to ensure comparison under a similar computational unit area, we use two chips for PBS and eight chips for NN. The results are shown in Table~\ref{tab:PBS_benchmarks} and Table~\ref{tab:NN_benchmarks}. 

\noindent \textbf{PBS~\cite{pbs}.} We evaluate the throughput of PBS, which is the main component in TFHE. We adopt the parameter set \uppercase\expandafter{\romannumeral 1} to \uppercase\expandafter{\romannumeral 4} in Strix~\cite{Strix}. 

\noindent \textbf{NN~\cite{NN}.} This involves a CNN inference on the MNIST dataset. The size of input images is 28 $\times$ 28. We evaluate the latency of NN with depths of 20, 50, and 100.


\subsubsection{Hybrid scheme benchmark}\label{sec:hybrid scheme benchmark}
We assess the performance of FlexMem on hybrid applications that incorporate both CKKS and TFHE. 

\noindent \textbf{HE$^3$DB~\cite{he3db}.} This is a homomorphic database evaluation that performs Query 6 in the TPC-H benchmark. The filter performs predicates based on TFHE, and the aggregation is performed based on CKKS. Scheme conversion~\cite{LWEs2RLWE} is performed between filter and aggregation. We evaluate the performance for 4096 and 16384 items, respectively. The results are shown in Table~\ref{tab:HE3DB_benchmarks}.

\begin{table}[]
    \centering
    \caption{Performance of conducting TPC-H query 6 (\textnormal{s}).}
    \label{tab:HE3DB_benchmarks}
    \resizebox{0.26\textwidth}{!}{
    \begin{tabular}{lrr}
        \toprule
        \textbf{Accelerators} & \textbf{4096} & \textbf{16384} \\ \hline
        CPU-baseline~\cite{he3db} & 3012 & 11835 \\
        Trinity~\cite{Trinity} & 0.42 & 1.68  \\
        \textbf{FlexMem} & \textbf{0.28} & \textbf{1.11} \\
        \bottomrule
    \end{tabular}}
\end{table}

\begin{figure}[t]
    \centering
    \includegraphics[scale=0.34]{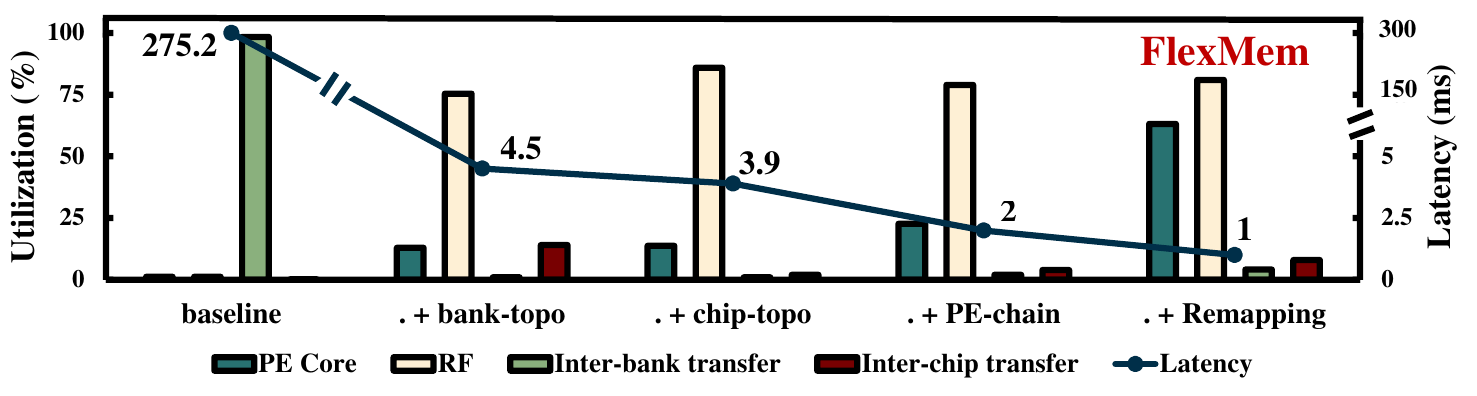}
    \caption{An ablation study of in-memory communication and runtime remapping designs in FlexMem. }
    \label{fig:ablation study}
    \vspace{-0.2cm}
\end{figure}

\subsection{Ablation study}\label{sec:ablation study}
To further analyze the effectiveness of our design in supporting the varying memory access strides. we conduct an ablation study on FlexMem, as illustrated in Fig.~\ref{fig:ablation study}. The baseline refers to FlexMem without PE chain, bank, or chip-level topology, and it employs a static data layout without remapping. In the absence of PE chains, data movement between PEs is handled via the global row buffer within the bank, following the process in Figaro~\cite{figaro}. Without a bank-level topology, data transfer is performed through the inter-bank bus as in RowClone~\cite{RowClone}. Besides, data transfer between chips is managed by the host CPU through the PCIe interface if without chip-level topology design. Each data group in Fig.~\ref{fig:ablation study} represents the previous group with additional optimization techniques. We observe that the bank and chip level topology designs significantly reduce the overhead caused by the memory access with strides across banks or chips. Furthermore, PE chain and remapping further improve the memory access performance of varying strides within a subarray row, leading to higher hardware utilization.

\begin{figure}[t]
    \centering
    \includegraphics[scale=0.35]{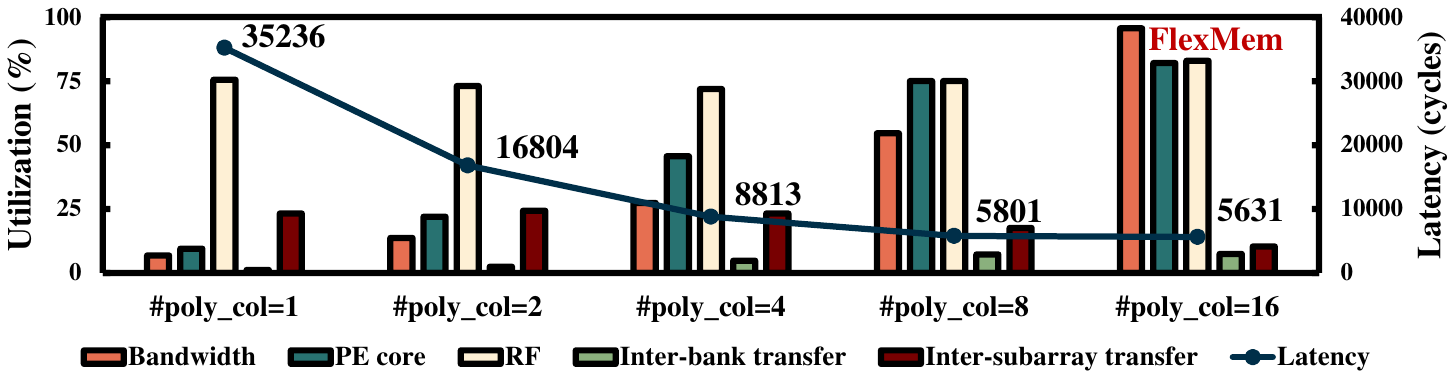}
    \caption{Latency and hardware utilization in 35 polynomials NTT with fixed $\#PE\_num=8$. }
    \label{fig:fixed_PE_num}
\end{figure}

\begin{figure}[]
    \centering
    \includegraphics[scale=0.35]{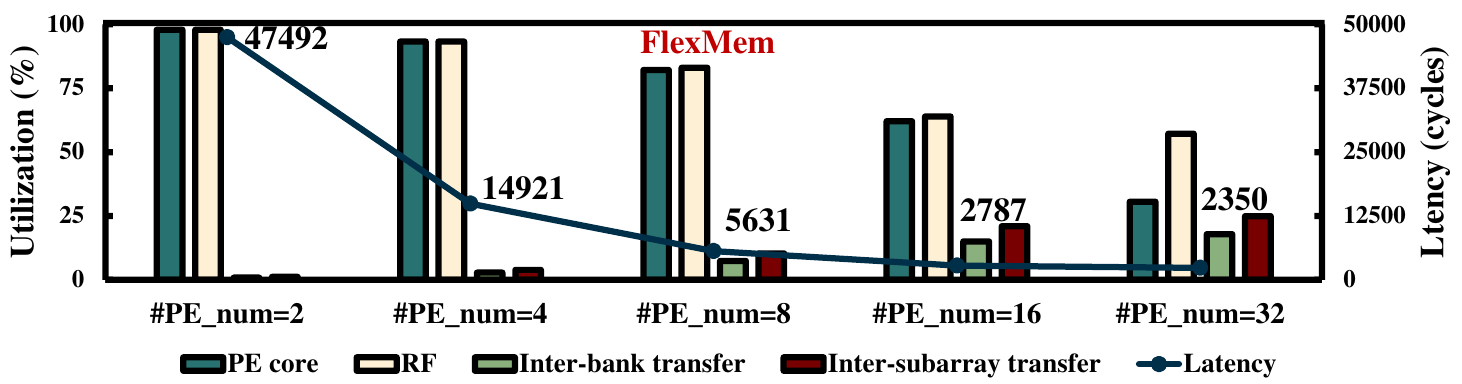}
    \caption{Latency and hardware utilization in 35 polynomials NTT with fixed $\#poly\_col =16$. }
    \label{fig:fixed_poly_col}
    \vspace{-0.2cm}
\end{figure}
\subsection{Design space exploration}\label{sec:design space exploration}
We explore two key parameters in FlexMem: $\#PE\_num$ and $\#poly\_col$, which represent the number of parallel PEs in each subarray pair and the number of columns occupied in a subarray per polynomial, respectively. First, we set $\#PE\_num = 8$, as in Fig.~\ref{fig:fixed_PE_num}. As the $\#poly\_col$ increases, the utilization of near-memory bandwidth, PE core, and RFs increases, leading to improvement that gradually saturates because of a bottleneck shifting from memory to computation. Similarly, as shown by Fig.~\ref{fig:fixed_poly_col}, when $\#poly\_col$ is fixed, and as $\#PE\_num$ increases, the bottleneck shifts from computation to memory. We select $\#PE\_num = 8$ in our design, as it achieves a balance between computational and memory access efficiency while maintaining low overhead and high performance. Besides, due to our remapping method, $\#poly\_col$ essentially remains at its achievable maximum value, which suggests a full in-memory bandwidth utilization at runtime.

\section{Conclusion}\label{sec:conclusion}
This paper proposes FlexMem, a DRAM-based NMP architecture for FHE. We analyze the limitations of traditional accelerators in off-chip memory access, as well as the low near-memory bandwidth utilization observed in NMP accelerators when dealing with varying memory access strides. To address these challenges, we design highly parallel computational units with hierarchical interconnection networks tailored to various memory access strides. Moreover, we propose dataflow designs at both polynomial and ciphertext levels, which enhance hardware utilization and parallel performance. Experiments demonstrate that FlexMem outperforms state-of-the-art traditional and NMP accelerators, achieving high acceleration performance and bandwidth utilization. 

\clearpage
\footnotesize
\bibliographystyle{IEEEtran}

\end{document}